\RequirePackage{fix-cm}
\documentclass[twocolumn,epjc3]{svjour3}
\smartqed  
\RequirePackage{graphicx}
\journalname{Eur. Phys. J. C}
\usepackage{amssymb}
\usepackage{graphics}
\usepackage{graphicx}
\usepackage{amsmath}
\usepackage{amsfonts}

\usepackage{xcolor}

\usepackage{bm}
\begin{document}

\title{Constraints on Tsallis Cosmology from Big Bang Nucleosynthesis \& Dark Matter Freeze-out
}
\subtitle{}


\author{Anish Ghoshal\thanksref{e1,addr1}
        \and
        Gaetano Lambiase\thanksref{e2,addr2,addr3}
}

\thankstext{e1}{e-mail: anish.ghoshal1@protonmail.ch}
\thankstext{e2}{e-mail: lambiase@sa.infn.it}


\institute{INFN, Rome Tor Vergata, Via della Ricerca Scientifica, I-00133 Rome, Italy \label{addr1} \and
 Dipartimento di Fisica E.R. Cainaiello, Universit\'a di Salerno, Via
Giovanni Paolo II, I 84084-Fisciano (SA), Italy \label{addr2}
            \and
            INFN, Gruppo Collegato di Salerno, Sezione di Napoli, Via Giovanni Paolo II, I 84084-Fisciano (SA), Italy \label{addr3}}

\date{Received: date / Accepted: date}

\maketitle

\begin{abstract}
We consider Tsallis cosmology as an approach to thermodynamic gravity and derive the bound on the Tsallis parameter to be $\beta<2$ by using the constraints derived from the formation of the primordial light elements, Helium, Deuterium and Litium, from the observational data from Big Bang Nucleosynthesis (BBN) which allows only a very tiny deviation from General Relativity (GR). Next we consider thermal dark matter (DM) freeze-out mechanism in Tsallis cosmological era and derive bounds on the Tsallis parameter from the observed DM relic abundance to be $1-\beta < 10^{-5}$.

\end{abstract}

\section{Introduction}
\label{intro}



Ever since the formulation of the laws of black hole mechanics was formulated by Bardeen et. al. \cite{BHLaws} and consequently the Bekenstein-Hawking entropy was understood \cite{Bekenstein,HawkingBHThermo}, gravity and entropy have been known to be very closely connected to each other. Moreoever recently the holographic principle \cite{Holography1,Holography2}, black hole complementarity \cite{Complementarity}, the gauge/gravity correspondence \cite{AdSCFT,Witten,MAGOO}, and the firewall puzzle \cite{AMPS,Braunstein} priciples and scenarios lead us to take seriously the relationship between gravitation and entropy and possibly hint towards the presence of such interconnection possibly even in the ultimate theory of quantum gravity. Classifying such approaches to gravity based on entanglement directly related, one may consider two distinct types, namely, \emph{holographic gravity} and \emph{thermodynamic gravity}.

In context to thermodynamic approach to gravity the seminal work by Jacobson \cite{Jacobson:1995ab}, showed that the Einstein field equations can be derived from the first law of
thermodynamics for the spacetime, many other approaches have been proposed \cite{Ver,Padmanabhan:2003gd,Padmanabhan:2009vy}.
One of the interesting consequences of the connection between gravity and thermodynamics is the derivation of
the cosmological equations, in particular the Friedman equations, obtained through the first law of thermodynamics
on the apparent horizon \cite{Elin,Cai1,Pad,Cai2,Cai3,CaiKim,Fro,verlinde,Cai4,CaiLM,Shey1,Shey2}.
To be mentioned that such an approach has been applied in differen models of modified gravity
\cite{Cai:2006rs,Akbar:2006er,Paranjape:2006ca,Sheykhi:2007zp,Jamil:2009eb,Cai:2009ph,Wang:2009zv,Jamil:2010di,Gim:2014nba,Fan:2014ala}.

All these studies rely on the standard thermodynamics. In the recent years,they have been generalized to the case
of extended entropy relations 
\cite{Tsallis:2012js,Komatsu:2013qia,Nunes:2014jra,Lymperis:2018iuz,Saridakis:2018unr,Sheykhi:2018dpn},
\cite{Artymowski:2018pyg,Abreu:2017hiy,Jawad:2018frc,Zadeh:2018wub,daSilva:2018ehn}.
Particular attention have been devoted to non-additive systems,
like gravitational system, in which the standard Boltzmann-Gibbs additive entropy turn out to be
generalized to the non-extensive Tsallis entropy \cite{Tsallis:1987eu,Lyra:1998wz,Wilk:1999dr}.

All these models have, as starting point, modifications of the area law of entropy.
with different motivations. For example, loop quantum gravity due to thermal
equilibrium fluctuations and quantum fluctuations
\cite{Log,Rovelli,Zhang}, or models of entropic cosmology which unifies the inflation and late
time acceleration \cite{YFCai} lead to logarithmic corrections, while
entanglement of quantum fields inside and outside the horizon
\cite{sau1,sau2,Sau,pavon1,lambiaseiorio} lead to modification for the area law
of entropy. The latter also arises in
systems with divergency in the partition function, like
gravitational system, so that the usual Boltzmann-Gibbs theory cannot be
applied, but instead one has to use the
generalized non-additive entropy \cite{Tsa0}. As a result, the microscopic mathematical expression of the
thermodynamical entropy of a black hole is given by \cite{Tsa},
\begin{eqnarray}\label{S}
S_{h}= \gamma A^{\beta}\,.
\end{eqnarray}
Here $A$ is the black hole horizon area, $\gamma$ a
constant and $\beta$ the Tsallis parameter or nonextensive
parameter \cite{Tsa}. The standard area law of
entropy is recovered as $\beta=1$ and $\gamma=1/(4L_p^2)$.

In the cosmological framework, the introduction of the Tsallis entropy provides
a modification of the Friedmann equations which contain extra terms, and reduces to the standard
cosmological model when the Tsallis generalized entropy becomes the usual one.

The aim of this paper is to explore the implications of the Tsallis cosmology on the formation
of light elements in the early Universe,  i.e. to the Big Bang Nucleosynthesis (BBN).
The latter occurred in the early phases of the Universe evolution, between the first fractions of  second after the Big Bang ($\sim
0.01$ sec) and a few hundreds of seconds after it, when the Universe  was hot and
dense (indeed BBN, together with cosmic microwave background radiation, provides the
strong evidence about the high temperatures characterizing the primordial Universe). It
describes the sequence of nuclear reactions that yielded the synthesis of light elements
\cite{kolb,bernstein,olivePDGroup2014}, and therefore drive the observed Universe. In general, from BBN
physics,  one may infer stringent constraints on a given cosmological model. In particular, in the present paper we
shall derive the  constraints on the free parameter $\beta$.

The layout of the paper is as follows. In the next Section we review the Tsallis cosmology. We discuss
the modified Friedmann equations obtained applying the first law of thermodynamics, $dE=T_hdS_h+WdV$,
at apparent horizon of a FRW Universe. To this aim, one uses Tsallis entropy associated with apparent horizon, see Eq. (\ref{S}).
In Section \ref{BBNanal} we use the BBN constraints to infer a bound on the free parameter of the Tsallis cosmology.
Conclusions are reported in Section \ref{Conclusions}.


\section{Tsallis cosmology from the First law of thermodynamics\label{FIRST}}

In this Section we shall recall the main features of the modification of the Friedman equations in the framework of the Tsallis statistics (here we closely follow the paper by Sheykhi \cite{Sheykhi}). As usual, we assume that, for a homogeneous and isotropic Universe (Friedamn-Robertson-Walker (FRW) Universe), the line element of the  given
\begin{equation}
ds^2={h}_{\mu \nu}dx^{\mu} dx^{\nu}+\tilde{r}^2(d\theta^2+\sin^2\theta d\phi^2),
\end{equation}
where $\tilde{r}=a(t)r$, $x^a=(t, r)$, $a=0, 1$, and $h_{\mu \nu}$=diag
$(-1, a^2/(1-kr^2))$ is the two dimensional metric. The spatial curvature $k$ refers to
open, flat, and closed Universes, corresponding to the values $k = 0, 1, -1$,
respectively. Moreover, for consistence with the laws of thermodynamics, it is assumed that the physical boundary of the
Universe is the apparent horizon with radius
\begin{equation}
\label{radius}
 \tilde{r}_A=\frac{1}{\sqrt{H^2+k/a^2}},
\end{equation}
to which is associated the temperature defined as $T_h=\frac{|\kappa|}{2\pi}=\Big|\frac{1}{2 \pi \tilde
r_A}\left(1-\frac{\dot {\tilde r}_A}{2H\tilde r_A}\right)\Big|$ (see Refs. \cite{Cai2,Sheykhi}).
Here $\kappa$ is the surface gravity and ${\dot r}\equiv dr/dt$. Usually it is assumed that  $\dot {\tilde r}_A\ll 2H\tilde r_A$, that means that the apparent horizon radius is fixed. This implies that the volume does not change and one may define\footnote{In Refs. \cite{cao} it has been studied the connection between temperature on the apparent horizon and the Hawking radiation, confirming the existence of the temperature associated with the apparent horizon.} $T=1/(2\pi \tilde r_A )$ \cite{CaiKim,Sheykhi}. The matter and energy content of the Universe is described by a perfect fluid, so that denoting with $\rho$ and $p$ the energy density and pressure, respectively, the corresponding stress-energy tensor reads
\begin{equation}\label{T1}
T_{\mu\nu}=(\rho+p)u_{\mu}u_{\nu}+pg_{\mu\nu},
\end{equation}
The conservation of the energy-stress tensor leads to the equation of continuity
\begin{equation}\label{Cont}
\dot{\rho}+3H(\rho+p)=0\,,
\end{equation}
where $H=\dot{a}/a$ is the Hubble parameter. One then define the {\it work density} as the work done by the volume change of the Universe, which is due to the change in the apparent horizon radius. For a FRW background, it is giving by $W=-\frac{1}{2} T^{\mu\nu}h_{\mu\nu}=\frac{1}{2}(\rho-p)$ \cite{Hay2,Sheykhi}.
This relation enters the first law of thermodynamics on the apparent horizon, which reads
\begin{equation}\label{FL}
dE = T_h dS_h + WdV\,.
\end{equation}
Using the fact that the total energy content of the universe inside a $3$-sphere of radius $\tilde{r}_{A}$ is $E=\rho V$ where
$V=\frac{4\pi}{3}\tilde{r}_{A}^{3}$ is the volume enveloped by 3-dimensional sphere with the area of apparent horizon
$A=4\pi\tilde{r}_{A}^{2}$, the equation of continuity (\ref{Cont}), and Tsallis entropy (\ref{S}), where now $A$ is the apparent horizon area of the Universe,
one infers the modified Friedmann equation obtained by means of the non-additive Tsallis entropy \cite{Sheykhi}
\begin{equation} \label{Fried4}
\left(H^2+\frac{k}{a^2}\right)^{2-\beta} = \frac{8\pi M_p^{-2\beta}} {3}
\rho,\end{equation}
provided we define
\begin{equation}\label{Lp}
\gamma\equiv\frac{3(2-\beta)(4\pi)^{1-\beta} M_p^{2\beta}}{4\beta } \,.
\end{equation}
From Eq. (\ref{Lp}) it follows that the non-additive parameter of the Tsallis entropy, the $\beta$-parameter, is always bounded, i.e.
\begin{equation}\label{betabound}
  \beta < 2
\end{equation}
The second modified Friedmann equation governing the evolution of the Universe in Tsallis cosmology (based
on the non-extensive Tsallis entropy) is given by \cite{Sheykhi}
\begin{eqnarray}
(4-2\beta) \frac{\ddot{a}}{a}+(2\beta-1)\left(H^2+\frac{k}{a^2}\right)^{2-\beta}=-\frac{8\pi}{M_{p}^{2\beta}} p\,. \label{2Fri3}
\end{eqnarray}
This equation can be derived by taking the first derivative with respect to $t$ of (\ref{Fried4}), using Eq. (\ref{Cont}) and  the relation
$\dot{H}=\ddot{a}/a-H^2$.
The combination of the first and second
modified Friedmann equations (\ref{Fried4}) and (\ref{2Fri3}) yields
\begin{eqnarray}\label{2Fri5}
\frac{\ddot{a}}{a}=-\frac{4\pi L_{p}^{2}\rho}{3(2-\beta)} \left[(2\beta-1) \rho +3p\right]\,.
\end{eqnarray}
Since the present Universe is currently in an acceleration expansion, hence $\ddot{a}>0$, one gets
\begin{eqnarray} \label{w1}
 (2\beta-1) \rho +3p <0  \  \     \longrightarrow   \  \  \omega< \frac{1-2\beta}{3},
\end{eqnarray}
where $\omega=p/\rho$ is the equation of state parameter.

Some comments are in order:

\begin{itemize}
  \item
  Cosmological observations coming from  Type Ia SNe \cite{Riess}, CMB\cite{Spergel} and the large scale structure \cite{Tegmark,Eisenstein}, provide the evidence that the current Universe is in an accelerated phase.To explain such observations, one has to introduce in cosmological models an exotic energy source (the Dark Energy (DE)) characterized by a negativepressure. At late times, the DE dominates over the cold dark matter, driving the Universe to the observed accelerating expansion. Remarkably, these observations can be directly derived in the Tsallis cosmological model. In fact, the inequality (\ref{w1}) implies that for $\beta\geq 1/2$ it follows $\omega< 0$, while for $\beta<1/2$ it follows $\omega\geq 0$. The latter result indicates that in Tsallis cosmology, the accelerated phase of the late time Universe is possible even with the ordinary matter (with positive equation of state parameter). Moreover, if $\omega=0$ (the Universe is now dominated with pressureless matter) then the accelerated expansion follows by  choosing the nonextensive parameter such that $\beta<1/2$.
Therefore, the accelerated expansion of the current Universe is possible without introducing any dark component.
   \item As shown in \cite{Sheykhi}, the generalized second law of thermodynamics in a region enclosed by the apparent horizon is still valid. In fact,
   denoting with $S_{h, m}$ the entropies associated to the horizon and matter, one can show, up to of irrelevant factor, that (see \cite{Sheykhi} for details)
\begin{equation}\label{GSL3}
T_{h}\frac{d}{dt}(S_{h}+S_{m})\sim \frac{1}{2-\beta} H {\tilde r_A}^{7-2\beta}(\rho+p)^2 \, > \, 0\,.
\end{equation}
For $\beta<2$, the right hand side of (\ref{GSL3}) is always a positive function, hence $\frac{d}{dt}(S_{h}+S_{m})\geq 0$. Thus the sum of the entropy of the boundary and the entropy of matter inside the bulk is a non decreasing function of time, according to the generalized second law of thermodynamics.
\end{itemize}


\subsection{Radiation-dominated era}\label{cosm}

We wish now analyze the Tsallis cosmology during the radiation dominated phase of the Universe expansion, which we are interested for the analysis of the primordial gravitational waves. For late convenience we shall rewrite (\ref{Fried4}) in the form
 \begin{equation}\label{H=AHGRIce}
   H(T)\equiv Z(T)\,H_\text{GR}(T)\,,
 \end{equation}
where $H_{GR}=\displaystyle{\sqrt{\frac{8\pi}{3M_{Pl}}\, \rho(T)}}$ is the Hubble parameter in the standard cosmology, and $A(T)$ is defined as
\begin{eqnarray}
  Z(T) &\equiv & \left[\sqrt{\frac{8\pi}{3}}\, \frac{\rho^{1/2}}{M_p^2}\right]^{\frac{\beta-1}{2-\beta}}
  \equiv   \eta \left(\frac{T}{M_{Pl}}\right)^\nu\,,  \label{ATsallis} \\
   \eta & \equiv & \left[\frac{2\pi}{3} \sqrt{\frac{\pi g_*(T)}{5}}\right]^{\frac{\beta-1}{2-\beta}}\,, \label{etaTsallis} \\
  \nu & \equiv & \frac{2(\beta-1)}{2-\beta} \label{nuTsallis}\,.
\end{eqnarray}
Here the relation\footnote{Notice that the Tsallis statistics affects also the definition of the thermodynamics quantities such as the number density, the energy density and the pressure. Typically one gets, in the relativistic regime,  $n\sim g_q T^3$, $\rho\sim g_q T^4$, $p=\rho/3$, where $g_q$ account for the $q$-corrections entering the distribution functions (the standard statistic is recovered in the limit $q\to 1$) \cite{lavagno}. Here we neglect such corrections.}
\begin{equation}\label{rhoTs}
\rho=\frac{\pi^2 g(T)}{30}\,T^4
\end{equation}
has been used, with $g(T)$ the effective number of degrees of freedom.  
For $\beta=1$, the parameter $\nu$ vanishes, $\nu=0$, so that the amplification factor $A(T)$ assumes the value $A(T)=1$, thus GR is recovered.

The equation of state for the radiation, $p=\rho/3$, implies that the continuity equation (\ref{Cont}) reads $\dot{\rho}(t)+4H\rho(t)=0$. A solution is $\rho(t)=\frac{\rho_0}{a^4(t)}$ \cite{Sheykhi}, where $\rho_0$ is a constant. Inserting this expression for $\rho$ in the Friedmann equation (\ref{Fried4}), one gets $a(t)= a_0 \left(\frac{2 t}{2-\beta}\right)^{1-\beta/2}$. Notice that the relation between the cosmic time $t$ and the temperature is $\frac{1}{t}\sim T^{\frac{2}{2-\beta}}$. From these relations one also infers $T a = constant$.



\section{Big Bang Nucleosynthesis  in Tsallis cosmology}
\label{BBNanal}

In this Section,  we examine the BBN in the framework of
Tsallis cosmology. The energy density of relativistic particles (hence we are considering the radiation
dominated era) filling up the Universe is given by Eq. (\ref{rhoTs}), with $g(T)\equiv g_*\sim 10$.
The neutron abundance is computed via the
conversion rate of protons into neutrons.  Here we recall the main features.

The formation of the primordial ${}^4 He$ occurs when
the temperature of the Universe was $T\sim 100$ MeV, while the
energy and number density were dominated by relativistic
leptons (electrons, positrons, neutrinos) and photons (the smattering of
neutrons and protons does not contribute in a relevant way to the
total energy density). These particles are in thermal
equilibrium owing to their rapid collisions, while the interactions of protons and
neutrons with leptons
\begin{eqnarray}
 \nu_e+n &\,\, \longleftrightarrow \,\, & p+e^- \label{int1} \\
 e^++n &\,\, \longleftrightarrow \,\, & p+{\bar \nu}_e \label{int2} \\
 n &\,\, \longleftrightarrow \,\, & p+e^- +{\bar \nu}_e \label{int3}
\end{eqnarray}
kept these particles in thermal equilibrium.
The neutron abundance in the expanding Universe is estimated by computing
the conversion rate of protons into neutrons, and its inverse, here indicated with $\lambda_{pn}(T)$ and $\lambda_{np}(T)$, respectively.
The weak interaction rate is (at enough high temperature)
\begin{equation}\label{Lambda}
\Lambda(T)=\lambda_{np}(T)+\lambda_{pn}(T)\,,
\end{equation}
in which $\lambda_{np}(T)$ is expressed as the sum of the rates associated
to the individual processes (\ref{int1})-(\ref{int3})
\begin{equation}\label{rate}
  \lambda_{np}=\lambda_{n+\nu_e \rightarrow  p+e^-}+
  \lambda_{n+e^+ \rightarrow  p+{\bar \nu}_e}+
  \lambda_{n \rightarrow  p+e^- +{\bar \nu}_e}\,.
\end{equation}
The quantities $\lambda_{np}$ and $\lambda_{pn}$ are related by $\lambda_{np}(T)=
e^{-Q/T}\lambda_{pn}(T)$, with $Q=m_n-m_p$ the difference of neutron and proton masses.
A comment is in order. During the freeze-out period, one may assume that (see \cite{bernstein}
for details) $a)$ the temperatures involved are the same, i.e. $T_\nu=T_e=T_\gamma=T$,
$b)$ $T<E$, i.e. the temperature $T$
is low with respect to the typical energies $E$ that contribute to
the integrals entering in the definition of the rates. The Fermi-Dirac distribution can be then replaced
with the Boltzmann distribution, $n_E \simeq e^{-E/T}$. $c)$ $m_e\ll E_e, E_\nu$, i.e. the electron mass
$m_e$ is negligible with respect to the electron and neutrino
energies.


Denoting with $ \Lambda$ the sum of the weak interaction rates,
 \[
 \Lambda\equiv\Lambda_{\nu_e+n  \leftrightarrow  p+e^-}+\Lambda_{e^++n  \leftrightarrow  p + {\bar \nu}_e}+
 \Lambda_{n \leftrightarrow  p+e^- + {\bar \nu}_e}
 \]
one obtains that the total rate reads \cite{bernstein,kolb}
 \begin{eqnarray}\label{Lambdafin}
    \Lambda({\cal T}) &=& 4 A\, {\cal T}^3(4! {\cal T}^2+2\times 3! {\cal Q}{\cal T}+2!
{\cal Q}^2) \\
       & \simeq & q T^5+{\cal O}\left(\frac{\cal Q}{T}\right) \nonumber
 \end{eqnarray}
where ${\cal Q}=m_n-m_p$ is the mass difference of neutron and proton,  $A=1.02 \times
10^{-11}$GeV$^{-4}$ and $q=9.6 \times 10^{-10}\text{GeV}^{-4}$. The primordial mass fraction of ${}^4 He$ can be estimated by making use of the relation \cite{kolb}
 \begin{equation}\label{Yp}
    Y_p\equiv \lambda \, \frac{2 x(t_f)}{1+x(t_f)}\,.
 \end{equation}
Here $\lambda=e^{-(t_n-t_f)/\tau}$, with $t_f$   the time of the freeze-out of the weak
interactions, $t_n$   the time of the freeze-out of the nucleosynthesis,
$\tau=8803\pm 1.1$sec the neutron mean lifetime \cite{olivePDGroup2014}, and $x(t_f)=e^{-{\cal
Q}/{\cal T}(t_f)}$ is the neutron-to-proton equilibrium ratio.
The function $\lambda(t_f)$ is interpreted as the fraction of neutrons that decay into
protons during the interval $t\in [t_f, t_n]$. Deviations from the fractional mass $Y_p$
due to the variation of the freezing temperature ${T}_f$ are given by
 \begin{equation}\label{deltaYp}
    \delta
Y_p=Y_p\left[\left(1-\frac{Y_p}{2\lambda}\right)\ln\left(\frac{2\lambda}{Y_p}
-1\right)-\frac{2t_f}{\tau}\right]
    \frac{\delta {T}_f}{{T}_f}\,,
 \end{equation}
where we have set $\delta {T}(t_n)=0$ since ${T}_n$ is fixed by the deuterium
binding energy \cite{torres,capozz}. A recent determination of mass fraction of ${}^4He$ has been obtained by using infrared and visible ${}^4He$ emission lines in 45 extragalactic HII regions. The analysis yields \cite{olive} (see also \cite{constraint})
 \begin{equation}\label{Ypvalues}
 Y_p=0.2449\pm 0.0040\,.
 \end{equation}
For our estimations on the Tsallis parameter $\beta$ we shall use (\ref{Ypvalues}) and therefore we shall take $|\delta Y_p| < 10^{-4}$.
Inserting these into (\ref{deltaYp}) one infers the upper bound
 \begin{equation}\label{deltaT/Tbound}
    \left|\frac{\delta {T}_f}{{T}_f}\right| < 4.7 \times 10^{-4}\,.
 \end{equation}
The relation $\Lambda= H$, where $\Lambda$ and $H$ are given by (\ref{Lambdafin}) and (\ref{H=AHGRIce}), respectively, allows to compute the freeze-out temperature
\begin{equation}\label{freezeout}
  T_f=M_{Pl} \left[\eta \frac{2\pi}{3} \sqrt{\frac{\pi g_*}{5}}  \frac{1}{q M_{Pl}^4}\right]^{\frac{1}{3-\nu}}\,.
\end{equation}
Defining $\delta T_f= T_f - T_{0f}$, with $T_{0f}\sim 0.6$ MeV (which follows from the standard computation with $H_{GR}\simeq q {T}^5$) one gets
\begin{equation}\label{Tffin}
 \left|\frac{\delta T_f}{T_{f}} \right|=\left|1-\frac{T_{0f}}{M_{Pl}}\left[\left(\frac{2\pi}{3} \sqrt{\frac{\pi g_*}{5}} \right)^{\frac{1}{2-\beta}}  \frac{1}{q M_{Pl}^4}\right]^{-\frac{1}{3-\nu}}\right| \,,
\end{equation}
with
 \[
 3-\nu=\frac{8-5\beta}{2-\beta}\,.
 \]
The constraint the free parameter $\beta$ is inferred using Eqs. (\ref{deltaT/Tbound}) and (\ref{Tffin}).

\begin{figure}[ht]
\includegraphics[width=3in,height=2.25in]{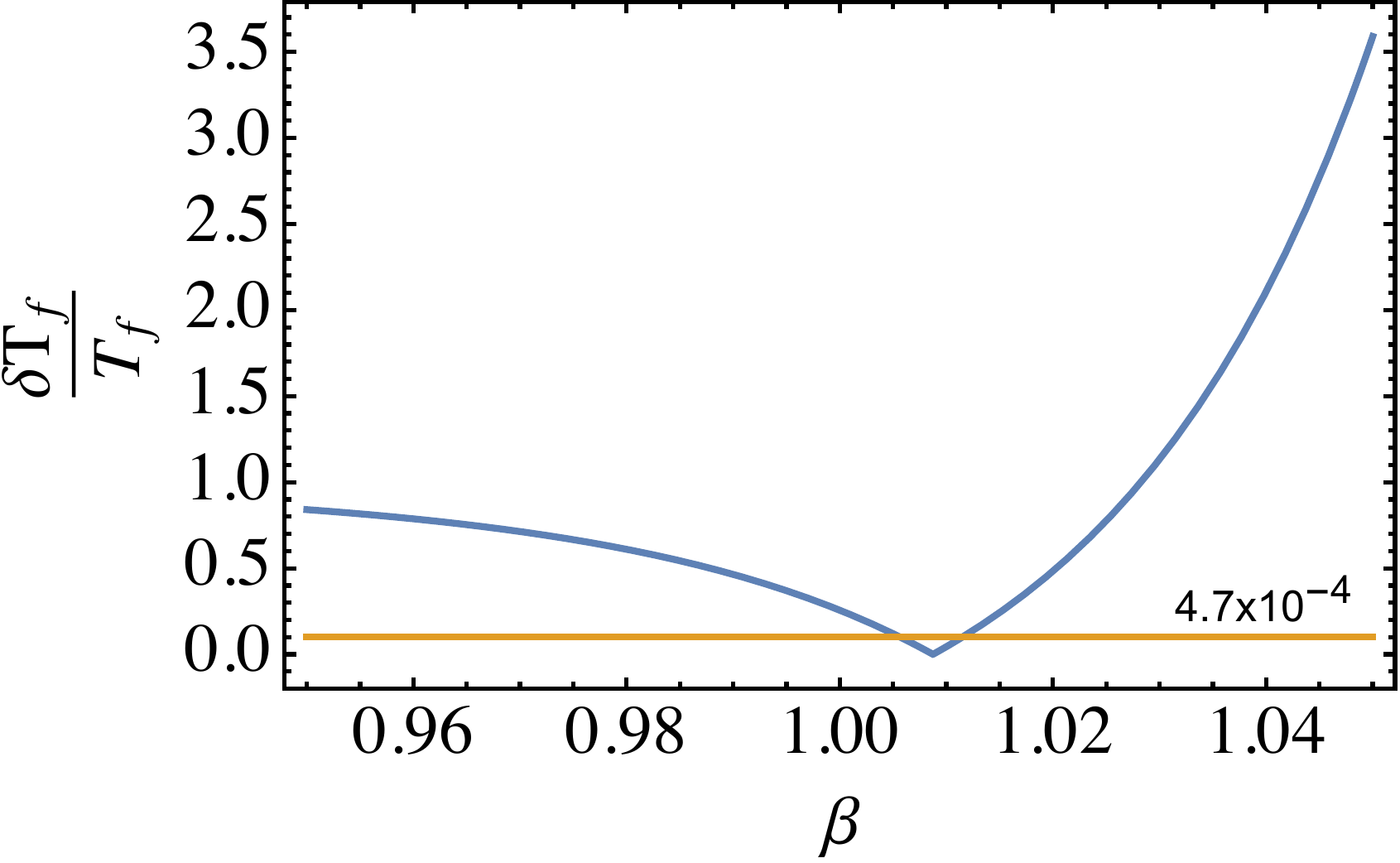}
\caption{\it $\delta {T}_{f}/T_{f}$ vs $\beta$. Here $\delta {T}_{f}/T_{f}$ is defined in (\ref{Tffin}), while the upper bound for $\delta {T}_f/{T}_{f}$ is given in Eq. (\ref{deltaT/Tbound}). The BBN provides the values $\beta \simeq 1.01$.}
\label{deltaTf1a}
\end{figure}

In Fig. \ref{deltaTf1a} we depict $\delta T_f/T_{f}$ from (\ref{Tffin})
vs $\beta$. The upper bound (\ref{deltaT/Tbound}) is also represented. As we can see,
constraints from BBN requires
 \begin{equation}\label{betabBBN}
  \beta\simeq 1.01
 \end{equation}
This results holds for early Universe.
It is consistent with the bound (\ref{betabound}), according to which the non-additive parameter $\beta$ of the Tsallis entropy
is always bounded. However, the constraint (\ref{betabBBN}) if applied to late Universe does not appear in agreement with
the bound $\beta<1/2$, needed to explain the current accelerated phase of the Universe.

\subsection{\bf Primordial light element $\{ {}^4 He$, $D, Li\}$ in Tsallis cosmology}

We will now derive the bound on the Tsallis parameter $\beta$ in a slightly different approach in which we analyze the effects of deviation from the standard cosmology on the primordial abundances of light elements, i.e. Deuterium  ${}^{2}H$, Helium ${}^{4}He$, and Tritium ${}^{7}Li$.
The basic idea is to replace the usual $Z$-factor entering the primordial light elements, which is related to the effective number of neutrinos species \cite{theory}
\[
Z_\nu = \left[ 1 + \frac{7}{43} (N_{\nu} - 3)\right]^{1/2}
\]
with the amplification factor entering (\ref{ATsallis}). In this analysis the baryon-antibaryon asymmetry, here indicated with $\eta_{10}$, plays a crucial role \cite{epjp52,epjp53}. Since we are interested to deviations from the standard cosmological model, hereafter we shall assume three generation of neutrinos so that we set $N_{\nu} = 3$. As we shall see, $Z\neq 1$ is hence provided by modification of GR, and in the specific case, by the Tsallis cosmology.

In what follows, we closely follow the work by Bhattacharjee abd Sahoo \cite{batt}. For completeness, we recall the main features:

\begin{itemize}

\item $^{4}He$ {\it abundance} - The production of Helium $^{4}He$ is generated by the production of $^{2}H$ through a neutron and a proton. Then, the Deuterium is converted into $^{3}He$ and Tritium ($T$). The relevant reactions are
\begin{eqnarray}
n + p &\rightarrow & {}^{2}H + \gamma; \\
{}^{2}H + {}^{2}H &\rightarrow & {}^{3}He + n;\\
{}^{2}H + {}^{2}H &\rightarrow & {}^{3}H + p
\end{eqnarray}
The Helium ${}^{4}He$ is finally produced owing to the following reactions 
\begin{equation}
{}^{2}H + {}^{3}H \rightarrow ^{4}He + n; \quad {}^{2}H + {}^{3}He\rightarrow {}^{4}He + p
\end{equation}
The numerical best fit provides the primordial ${}^{4}He$ abundance \cite{epjp57,epjp58}
\begin{eqnarray}
Y_{p} &=& 0.2485 \pm 0.0006 + \\
       & & +  0.0016 \left[\left( \eta_{ 10} - 6\right) +100\left( Z-1\right)  \right]\,, \nonumber
\end{eqnarray}
where $Z$ is now  given by (\ref{ATsallis}), while the baryon density parameter $\eta_{10}$ is defined as \cite{epjp52,epjp53}
\begin{equation}\label{eta10}
\eta_{10} \equiv 10^{10}\eta_{B}\equiv 10^{10} \frac{\eta_{B}}{\eta_{\gamma}}\,,
\end{equation}
with $\eta_{10} \simeq 6 $. Here $\eta_{B}=n_B/n_\gamma$ represents the baryon to photon ratio \cite{epjp56}.
For $Z=1$, the standard result of BBN for the $^{4}He$ fraction is recovered, so that $(Y_{p})|_{GR} = 0.2485 \pm 0.0006$.
The observational data for the Helium $^{4}He$ and setting $\eta_{10} = 6$ imply that
the abundance is $0.2449 \pm 0.0040$ \cite{jcap2020}, so that
\begin{eqnarray}
0.2449 \pm 0.0040 &=& 0.2485 \pm 0.0006 + \\
       & & + 0.0016 \left[100(Z-1) \right]\,. \nonumber
\end{eqnarray}
From here one infers the constrain
\begin{equation}\label{ZHe4}
  Z = 1.0475 \pm 0.105\,.
\end{equation}

\item $^{2}H$ {\it abundance} -  Deuterium $^{2}H$ is produced from the reaction $n+p \rightarrow {}^{2}H +\gamma$.
The numerical best fit provides the following Deuterium abundance \cite{epjp52}
\begin{equation}\label{16}
y_{D p} = 2.6(1 \pm 0.06) \left( \frac{6}{\eta_{ 10} - 6 (Z-1)}\right)^{1.6}
\end{equation}
The values $Z=1$ and $\eta_{10} = 6$ yields the standard result in GR $y_{D p} |_{GR} = 2.6 \pm 0.16$.
Equating the observational constraint on deuterium abundance $y_{D p} = 2.55 \pm 0.03$ \cite{jcap2020} with Eq. (\ref{16}) one gets
\begin{equation}
2.55 \pm 0.03 = 2.6(1 \pm 0.06) \left( \frac{6}{\eta_{ 10} - 6 (Z-1)}\right)^{1.6}
\end{equation}
The constraint on $Z$ is
\begin{equation}\label{ZD}
Z =1.062 \pm 0.444\,.
\end{equation}
This constraint partially overlaps with that of the helium abundance (\ref{ZHe4}).

\item $^{7}Li$ {\it abundance} - As regards the Lithium abundance, there is, as well known, a puzzle (referred in literature as {\it the Lithium problem} \cite{theory}) since the $\eta_{10}$ parameter defined in (\ref{eta10}) from a side successfully fits the abundances of $D$ and ${}^4He$, from the other does not fit the observations of $^{7}Li$. The ratio of the expected value of ${}^{7}Li$ abundance in the standard cosmological model with respect to the observed one is in the range \cite{theory,theory42}
    \begin{equation}\label{LitabGR}
    \frac{Li|_{GR}}{Li|_{obs}}  \in [2.4-4.3]\,.
    \end{equation}
Therefore neither the standard BBN nor any modified cosmology allow to low abundance of $^{7}Li$. The numerical best fit for $^{7}Li$ abundance is \cite{epjp52}
\begin{equation}
y_{Li} = 4.82 (1 \pm 0.1)\left[\frac{\eta_{ 10} - 3 (Z-1)}{6} \right] ^{2}
\end{equation}
Observational constraint on Lithium abundance is $y_{Li} = 1.6 \pm 0.3$ \cite{jcap2020}. This leads to
the constraint on $Z$ 
\begin{equation}\label{ZLi}
Z = 1.960025 \pm 0.076675\,.
\end{equation}
Such a value does not overlap with the constraints on ${}^2 H$ abundance, Eq. (\ref{ZD}), and on ${}^4He$ abundance, Eq. (\ref{ZHe4}).

\end{itemize}

\section*{Discussion of the obtained results}

\begin{figure}[ht]
  \includegraphics[width=3in,height=2.25in]{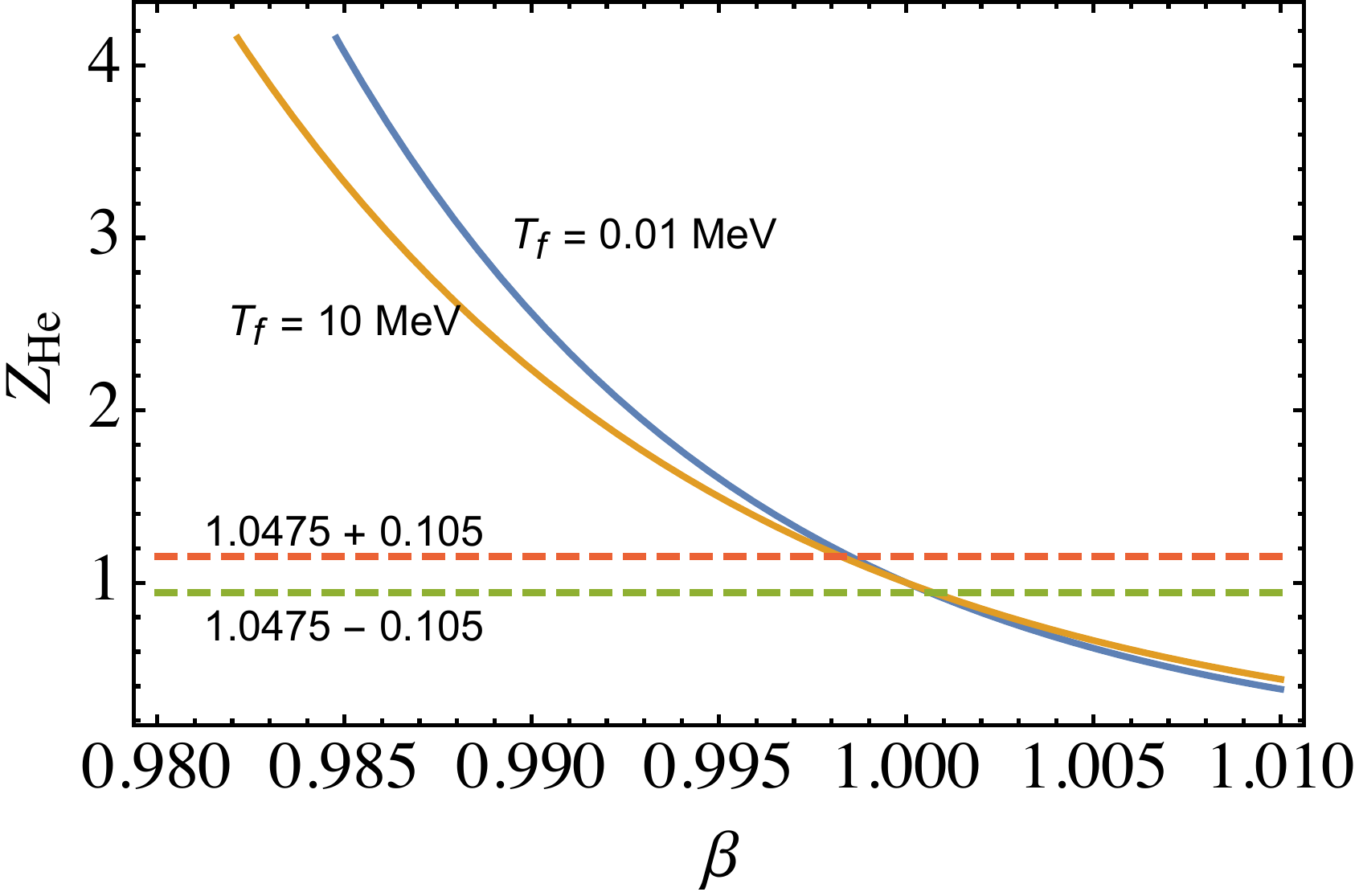}
  \caption{$Z_{{}^4 He}$ vs $\beta$. The experimental range (\ref{ZHe4}) is reported. We have fixed the baryon parameter to $\eta_{10}=6$ and varied the freeze-out temperature in the range  $T_f= [0.01, 10]$MeV.}
  \label{Helium}
\end{figure}

\begin{figure}[ht]
  \includegraphics[width=3in,height=2.25in]{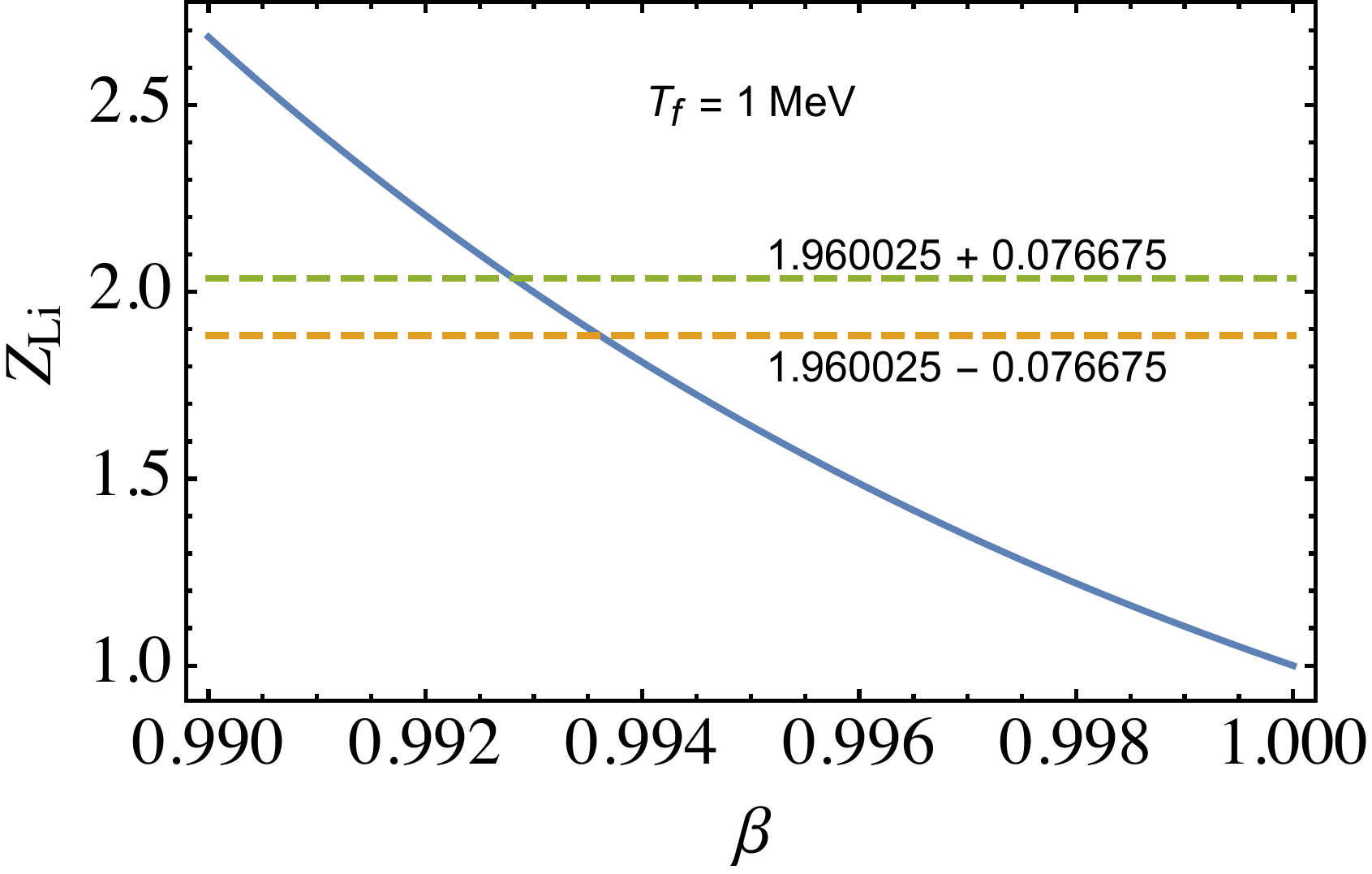}
  \caption{$Z_{Li}$ vs $\beta$. The experimental range (\ref{ZLi}) is reported. We have fixed the baryon parameter to $\eta_{10}=6$ and the freeze-out temperature to $T_f= 1$ MeV.}
  \label{Litium}
\end{figure}

Here we discuss the previous results, deriving the bound on the Tsallis parameter $\beta$. In Fig. \ref{Helium} we plot (\ref{ATsallis}) taking into account the constraints (\ref{ZHe4}). As we can see, the corresponding range of variability of $\beta$ is
 \begin{equation}\label{betaboundHe4}
 0.996 \lesssim \beta \lesssim 1.001\,.
 \end{equation}
Similar results occur for the Deuterium (using Eq. (\ref{ZD})). In the case of Litium, the plot \ref{Litium} indicates that, using the constraints provides by Eq. (\ref{ZLi}), the parameter $\beta$ varies in the range
 \begin{equation}\label{betaboundLi}
 0.993\lesssim \beta \lesssim 0.994\,.
 \end{equation}
Although there is not an overlapping of the two range, (\ref{betaboundHe4}) and (\ref{betaboundLi}), their difference is very close, suggesting the possibility that the Litium problem could be solved in the framework of modified cosmology.

\section{Tsallis cosmology as an effective $f(R)$ cosmology and bounds from Dark Matter}


In this Section we derive the bound on $\beta$ by making use of Dark Matter (DM) Annihilation Cross Section entering the cold DM relic abundance $\Omega_{cdm}$
To this aim, first we show that Tsallis cosmology is related to modified cosmology described by $f(R)=\alpha R^n$ model. In such a way, we can borrow results of Ref to connect the Tsallis parameters with cold DM abundance.

The field equations for $f(R)$ theory are inferred by varying the action $S=\int d^4x \sqrt{-g}f(R)$ with respect to the tensor metric $g_{\mu\nu}$
\begin{equation}\label{fieldeqs}
  f' R_{\mu\nu}-\frac{f}{2}\, g_{\mu\nu}-\nabla_\mu \nabla_\nu f' +g_{\mu\nu}\Box f' = \frac{8\pi}{3M_{Pl}^2} T^m_{\mu\nu}\,,
\end{equation}
where $f^\prime\equiv \displaystyle{\frac{\partial f}{\partial R}}$, and $T^m_{\mu\nu}$ is, as usual, the energy momentum tensor for matter, which fulfill the Bianchi identities $\nabla_\mu  T^{m\,\mu\nu}=0$. The trace of (\ref{fieldeqs}) gives
\begin{equation}\label{tracef}
  3\Box f'+f' R-2f=\kappa^2 T^m\,, \qquad T^m = \rho-3p\,.
\end{equation}
In a (spatially flat) FRW metric, and assuming that the scale factor is of the form $a(t)=a_0 t^\gamma$, Eqs. (\ref{fieldeqs}) and (\ref{tracef}) read
 \begin{eqnarray}\label{Hmodified}
   \alpha \Omega R^n &=& \kappa^2 \rho\,, \\
   \alpha \Gamma R^n &=& \kappa^2 T^m \,, \label{Hmodifiedtrace}
 \end{eqnarray}
where
 \begin{eqnarray}
   \Omega & \equiv &  \frac{1}{2}\left[\frac{n(\gamma+2n-3)}{2\gamma-1}-1\right]\,, \label{Omega}\\
   \Gamma &\equiv & n-2-\frac{n(n-1)(2n-1)}{\gamma(2\gamma-1)}+\frac{3n(n-1)}{2\gamma-1}\,, \label{Gamma} \\
   R &=& \frac{6\beta(1-2\gamma)}{t^2}\,. \label{Rdef}
 \end{eqnarray}
and we have used $\Box f'= {\ddot f}^\prime+3H{\dot f}^{\prime}$. The combination of Eqs. (\ref{Omega}) and (\ref{Rdef}) allows to recast the FRW equation (\ref{Hmodified}) in the form (\ref{H=AHGRIce}), i.e.
 \begin{eqnarray}\label{ZfR}
   Z_{f(R)} &=& \eta_{f(R)}\left(\frac{T}{M_{Pl}}\right)^{\nu_{f(R)}}\,, \\
   \eta_{f(R)} &=& \sqrt{\frac{\gamma (8\pi/3)^{(1-n)/n}}{6|2\gamma-1|}} \left(\frac{\pi^2 g_*}{30}\right)^{\frac{1-n}{2n}}
   \frac{1}{({\tilde \alpha}\Omega)^{1/2n}}\,,  \label{etafR}\\
   \nu_{f(R)} &=& \frac{2}{n}-2\,, \label{nufR} 
 \end{eqnarray}
where the adimensional constant ${\tilde \alpha}$ is related to the constant $\alpha$ by the relation $ {\tilde \alpha}={\alpha}{M_{Pl}^{-2(1-n)}}$.

The comparison of (\ref{nuTsallis}) and (\ref{nufR}) gives the relation between $\beta$ and $n$,
 \begin{equation}\label{nbeta}
   n-1=1-\beta\,,
 \end{equation}
while the comparison of (\ref{etaTsallis}) and (\ref{etafR}) provides the expression of the constant $\alpha$ in terms of $\beta$ and $\gamma$, needed for
interpreting the Tsallis cosmology as an effective $\alpha R^n$ cosmology. Since this expression is not relevant for what follows, we shall not write it explicitly.

Next we discuss the previous results in the framework of DM relic abundance, assuming that DM is composed of weakly-interacting massive particles (WIMPs).
Following \cite{cross}, the DM relic density in modified cosmology is given by
\begin{equation}
\Omega_{cdm}h^{2} \simeq 10^{9} \frac{(\bar{l} + 1) x_{f}^{(\bar{l} + 1)} \text{GeV}^{-1}}{(h_{*}/g_{*s}^{1/2}) M_{Pl}\bar{\sigma}}\,,
\end{equation}
where $x_{f}$ is the freeze-out temperature \cite{turner,cross}
\begin{eqnarray}\label{xfDM}
x_{f} &=& \ln [0.038 (\bar{l} + 1) (g / g_{*s}^{1/2}) M_{p} m \bar{\sigma}] - \\
 &-& (\bar{l} + 1) \ln[\ln [0.038 (\bar{l} + 1) (g / g_{*s}^{1/2}) M_{p} m \bar{\sigma}]] \nonumber
\end{eqnarray}
with $g=2$ the spin polarizations of the dark matter particle and $m$ the mass of WIMPs particles, $\bar{\sigma}$ the WIMP cross section, and
\begin{equation}\label{barl}
\bar{l} = l + \left(\beta-1 \right)\,.
\end{equation}
Here\footnote{We note that in \cite{cross} it has been used the parametrization $\langle \sigma v\rangle = \sigma_0 x^{-l}$, where $l=0$ corresponds to $s$-wave annihilation, $l=1$ to $p$-wave
annihilation, and so on. The modification of standard cosmology induces the corrections to the parameter $l$, see Eq. (\ref{barl}). When $\beta=1$, hence the evolution of the Universe is described by the standard cosmological model, Eq. (\ref{barl}) implies ${\bar l}=l$, reproducing the standard results.} $\bar{l} = l$ ($\beta = 1$) for GR, while $l=0, 1$ correspond to $s$-wave and $p$-wave polarizations, respectively.
Recent cosmological observations have constrained the normalized cold dark matter density in the range \cite{cross18}
\begin{equation}
0.075 \lesssim \Omega_{cdm}h^{2} \lesssim 0.126
\end{equation}
Following the analysis of \cite{cross}, one infers
\begin{equation}\label{boundbetaDM}
  1-\beta \lesssim 0.00016\,,
\end{equation}
that is DM relic density provides a tiny deviation of Tsallis cosmology from the standard cosmological model, although such a deviation could be non negligible for the WIMPS cross section (see \cite{cross} for details).


\section{Conclusions}
\label{Conclusions}

In this paper we have investigated the consequences of Tsallis cosmology on the formation of primordial light elements.  We summarize the findings of our paper as following:
\begin{enumerate}
    \item We have shown that the viable Tsallis cosmology, characterized by the bound on the Tsallis parameter $\beta<2$, satisfies the BBN constraints.
More precisely, we have shown that the BBN constraint induces a value for the free parameter of the Tsallis cosmology of the order
$\beta \simeq {\cal O}(1)$. The analysis of the light elements ${}^4 He, {}^2 H, Li$ indicates that the range of variability of $\beta$, inferred taking into account the bounds (\ref{ZHe4}), (\ref{ZD}) and (\ref{ZLi}), does not overlap, although are very close. This suggest the possibility to search for cosmological models in order to solve the Litium puzzle.
\item We have also shown that the Tsallis cosmology can be seen as an effective modified cosmology. In particular we have shown that the $f(R)$ model, with $f(R)=\alpha R^n$, leads to the same Friedman field equations of the Tsallis cosmology provided that the parameters characterizing the two models are related as $n-1=1-\beta$, see Eq. (\ref{nbeta}). Such a result has been then used to constraint $\beta$ by using the observational data on DM relic abundance, $1-\beta < 10^{-5}$, see Eq. (\ref{boundbetaDM}).
\item It must be pointed out, however, that the value of $\beta\sim {\cal O}(1)$ here derived in the framework of BBN is not compatible with the bound $\beta < 0.5$, needed to explain the observed accelerating phase of the present Universe, unless some mechanism occured during the evolution of the Universe able to reduce the value of the parameter $\beta \sim {\cal O}(1)$ to values $\beta<0.5$. This could be the case recently discussed by Nojiri, Odintsov and Saridakis \cite{nojirivar}, which have proposed  a modified cosmological scenario arising from the application of non-extensive thermodynamics with varying Tsallis parameter.

\item In conclusions, the results here discussed could contribute in the debate of fixing the most realistic scenario among models based on Tsallis cosmology (hence the entropic origin of gravity) and dark sector. Much work is needed in this direction, and will be faced elsewhere.
\end{enumerate}

In future following our work, bounds on Tsallis cosmology can derived from observed baryon asymmetry
in the universe similar to the baryon asymmetry forming in Tsallis cosmological era. Moreover, Primordial Gravitational Wave (GW) signals can be looked into if the tensor perturbations which originated during inflation propagated during Tsallis cosmological era and possibility of detecting in current and future GW detectors following our earlier work in Ref. \cite{Bernal:2020ywq}.

\begin{acknowledgements}
G.L. thanks INFN and MIUR for support.
\end{acknowledgements}



\end{document}